\documentclass[aps,prb,showpacs,twocolumn,floats,pdffig,pdflatex,floatfix]{revtex4}
\usepackage{amssymb}
\usepackage{amsbsy}
\usepackage{amsmath}
\usepackage{epsfig}
\usepackage{graphicx}
\usepackage{color}

\newcommand\beq{\begin{equation}}
\newcommand\eeq{\end{equation}}
\newcommand\bea{\begin{eqnarray}}
\newcommand\eea{\end{eqnarray}}

\begin{document}

\title {Stationary Charge Imbalance Effect in System of Coupled Josephson Junction}

\author{K. V. Kulikov$^1$, M. Nashaat$^{1,2}$ and Yu. M. Shukrinov$^{1,3}$}

\affiliation{
	$^1$ BLTP, JINR, Dubna, Moscow region, 141980, Russia. \\
	$^2$ Department of Physics, Cairo University, Cairo, Egypt\\
	$^3$ Dubna State University, Dubna, Russia}
\date{\today}

\begin{abstract}
We investigate stationary charge imbalance effect in the system of coupled overdamped Josephson junctions. We show that coupling between junction and nonzero stationary charge imbalance in the resistive state bring to a decrease of the Josephson frequency in the Josephson junctions of the stack. The formed difference in Josephson frequency leads to the nonuniform switch to the Shapiro step regime in the presence of external electromagnetic radiation and appearance of kinks of voltage on the IV-characteristics of the stack. We also show that stationary charge imbalance brings to the slope of the Shapiro step due to the difference of the charge imbalance potential on the edges of the step. The theoretical and experimental results for voltage bias coupled Josephson junctions have been compared with the current bias case.
\end{abstract}



\maketitle

\section{Introduction}

The phase dynamics of the layered superconducting materials have attracted a great interest because of rich and interesting physics from one side and perspective of applications from the other one \cite{bkg_prb-11, k_prb-10, pasu_prb-08, yurgens2000intrinsic, okk_sci-07}. In particular, the nonequilibrium effects created by stationary current injection in high-$T_c$ materials have been studied very intensively in recent years \cite{Artemenko97,Preis98,Shafranjuk99, Helm00,Helm01,Helm02,Bulaevskii02}. However,  the charge imbalance in the systematic perturbation theory is considered only indirectly as far as it is induced by fluctuations of the scalar potential \cite{Artemenko97, Preis98,Helm01}. In Ref.\cite{keller05} it is taken into account as an independent degree of freedom and, therefore, the results are different from those of earlier treatments. In addition, due to the fact that charge does not screen in the superconducting layers, a system form a coupled Josephson junctions (JJs) \cite{Koyama96,Dmitry}. 


Last few years two theoretical models are widely used to describe intrinsic Josephson junctions: capacitively coupled Josephson junctions (CCJJ) model \cite{koyama1996v}, based on quasi-neutrality breakdown effect which is appears due to the comparable thickness of the superconducting layers with the Debye length and charge imbalance (CIB) model \cite{epl, sm-prl2007, keller05, rother03, Ryndy-kkell, Dmitry, ryndyk1997quasiparticle}, based on the quasi-particle charge imbalance effect induced by normal current injection. Actually, relaxation length of charge imbalance in the layered system could be much larger than any other characteristic lengths. Therefore, it could exist in HTSC because the thickness of superconducting layers is smaller than the Debye screening length and thus, obviously less than characteristic length of nonquilibrium relaxation.

Experimentally for low temperature superconductors the data yielded in Ref's. \cite{clarke1972experimental, yu1972electric, clarke1974measurements} provided a strong evidence for charge imbalance effect. In these experiments a difference between the quasi-particle potential and Cooper pair chemical potential in the nonequilibrium region is detected. In Ref. \cite{rother03} the voltage biased experiments are done for mesas structure of BSCCO. In these experiments two important effects are observed for stationary case and they are explained as a result of charge imbalance on the superconductor layers. The first effect is the shifting of the Shapiro step from its canonical value $\hbar \omega_R/(2e)$. While, the second one, an influence of the current through one mesa on the voltage drop across the other mesa, is occurred when two mesas structure close to each other on the same base crystal.

In Ref.\cite{epl} we have shown that in the system of intrinsic Josephson junctions of high temperature superconductors under external electromagnetic radiation the charge imbalance is responsible for a slope in the Shapiro step in the IV-characteristic. We found that the value of slope increases with a nonequilibrium parameter, coupling between junctions leads to the distribution of the slope's values along the stack and the nonperiodic boundary conditions shift the Shapiro step from its canonical position.

In this study, we consider the current biased coupled overdamped Josephson junctions in the presence of stationary charge imbalance. The IV-characteristics of coupled JJ are numerically calculated using the resistively shunted junction model. Our model takes into account both coupling between the layers and quasiparticle charge imbalance effect \cite{epl, sm-prl2007, keller05, rother03, Ryndy-kkell, Dmitry, ryndyk1997quasiparticle}. We show that stationary charge imbalance leads to a change in Josephson frequency value in JJs of the stack. The difference in frequency along the stack brings to the appearance of kinks on the IV-characteristics below and above Shapiro step. We also demonstrate a slope and shift of the step caused by the stationary charge imbalance.

\section{Model}
The key point of the theory developed in \cite{Artemenko97, Koyama96, Dmitry} is the nonequilibrium character of the Josephson effect in layered superconductors. Superconducting layers with a thickness $d_s^l$ less than the characteristic length of the nonequilibrium relaxation $l_E$ and the Debye screening length $r_{D}$ are in a non-stationary nonequilibrium state due to the injection of quasiparticles and Cooper pairs, and a non-zero invariant potential formed inside them

\begin{eqnarray}
 	\Phi_l(t)= \phi_l + \frac{\hbar}{2e}\dot{\theta}_{l},
\end{eqnarray}

where $\phi_l$ is the electrostatic potential, $\theta$ is the phase of superconducting condensate, $\Phi=0$ in the equilibrium state (here and below $e=\mid e \mid$). It is important to note that in the non-equilibrium mode, the usual Josephson ratio $d\varphi_l/dt = (\hbar/2e)V$, which connects the Josephson phase difference $\varphi_{l}(t)$ between the layers $l-1$ and $l$, and the voltage $V_l=\phi_{l-1}-\phi_l$ \cite{Artemenko97, Koyama96, Dmitry} is violated and instead a generalized Josephson relation appears

\begin{eqnarray}
 	\frac{d \varphi_l}{dt}= \frac{2e}{\hbar}V_l + \frac{2e}{\hbar}(\Phi_{l-1}-\Phi_l).
\end{eqnarray}

Thus, $\Phi_l$ are new important dynamic variables of the theory. Note that the shift of the chemical potential of a superconducting condensate from its equilibrium value is $\delta \mu_s =e \Phi$ and is determined from the expression for the charge density inside the superconducting layer \cite{av_spu-79, e-ZETF-71}

\begin{eqnarray}
 	\rho=-2e^2 N(0)(\Phi_l - \Psi_l)= -\frac{1}{4\pi r^2_D}(\Phi_l - \Psi_l),
\end{eqnarray}

where $\Psi_l$ is the potential determined by the electron-hole imbalance of the branches of the elementary excitations of superconductor and $N(0)$ is the density of states.

\begin{figure}[htb]
 \centering
\includegraphics[width=80mm]{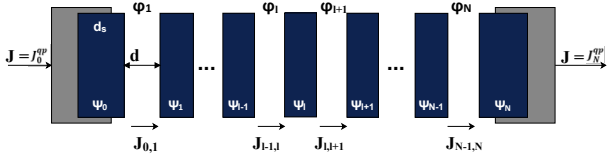}
\caption{Layered system of $N+1$ superconducting layers forms a stack of Josephson junctions. Since the 0-th and N-th layers are in contact with normal metal, their thicknesses $d^0_s$ and $d^N_s$  are different from the thickness of the other S-layers $d_s$ inside of the stack due to the proximity effect.}
 \label{fig:layer}
\end{figure}

A stack of $N+1$ superconducting layers (S-layers) forming a system of $N$ coupled JJs is presented in Fig. \ref{fig:layer}. 
The Josephson relation can be rewritten, taking into account the coupling between the S-layers and stationary charge imbalance potential, in the form \cite{sm-physc2006, SM-PhysC2006}

\begin{eqnarray}
\frac{d\varphi_l(t)}{dt} &=& \frac{2e}{\hbar} \Bigl( (1+2 \alpha ) V_{l}(t)-\alpha (V_{l-1}(t) + V_{l+1}(t)) \\ &+& \Psi_{l}(t) - \Psi_{l-1}(t) \Bigr),
\end{eqnarray}

with a voltage $V_l$ between the layers $l-1$ and $l$, $ V_l(t) \equiv V_{l,l-1}(t) $, $\varphi_{l}(t)$ is the phase difference across the layers $l-1$ and $l$, $\Psi_{l}(t)$ is the charge imbalance potential on the layer $l$ and $\alpha = \epsilon\epsilon_{o} /2e^{2} N(0) d_s d_i $ is the coupling parameter, $\epsilon$ is the dielectric constant, $\epsilon_{o}$ is the vacuum permittivity, $d_s$ is thickness of superconducting layers,  $d_i$ is thickness of insulating layers, and $N(0)$ is the density of states.

The total current density $J_{l-1,l} \equiv J_{l}$ through each S-layer is given as a sum of superconducting, quasiparticle and diffusion terms:

\begin{eqnarray}
 	J_{l}=J_{c}\sin\varphi_{l}+\frac{\hbar}{2eR}\dot{\varphi}_{l}+\frac{\Psi_{l-1}-\Psi_{l}}{R},
\end{eqnarray}

where $J_{c}$ is the critical current density, and $R$ is the junction resistance. This equation together with the generalized Josephson relation and kinetic equations for $\Psi_l$

\begin{eqnarray}\label{Psi_poten}
 	\Psi_l= \eta (J_{l}^{qp}-J_{l-1}^{qp})
\end{eqnarray}

describe the dynamics of coupled JJs. In formula (\ref{Psi_poten}) $\eta=\frac{4\pi r_D^2\tau_{qp}}{d^{l}_{s} d_i R}$ is nonequilibrium parameter, $\tau_{qp}$ is the quasiparticle relaxation time. We use normalized equation for current to calculate phase dynamics  

\begin{eqnarray}
 	\dot{\varphi}_{l} &=& I -\sin\varphi_{l} + A \sin{\omega} \tau + I_{noise} +\psi_{l}-\psi_{l-1},
 	\label{eq:1}
\end{eqnarray}  

where the dot shows a derivative with respect to $\tau=\omega_ct$, $I=J/J_{c}$ is the dimensionless current density, $\omega_c=\frac{2eJ_cR}{\hbar}$ is characteristic frequency. This equation is solved numerically using the fourth order Runge-Kutta method. In parallel, at each time step we use normalized kinetic equations with nonperiodic boundary conditions to calculate nonequilibrium potential by means of Gauss method 

\begin{eqnarray}
 	(1+\eta)\psi_0 - \eta \psi_1 &=& \eta (I + A\sin{\omega} \tau - \dot{\varphi}_1)\nonumber \\
 	-\eta \psi_{i-1} + (1+2\eta)\psi_{i} - \eta \psi_{i+1} &=&  \eta (\dot{\varphi}_i - \dot{\varphi}_{i+1})\\
 	\eta \psi_{N-1} + (1-\eta)\psi_N &=&  \eta(\dot{\varphi}_N - I -  A\sin{\omega \tau})\nonumber \hspace{0.3cm}
 	\label{eq:2}
 \end{eqnarray}

Then we calculate voltage across the JJ by means of normalized Josephson relation with non periodic boundary condition
 
\begin{eqnarray}
 	\dot{\varphi}_{1} &=& v_{1} - \alpha (v_{2}-(1+\gamma)v_{1}) +\psi_{1}-\psi_{0}\nonumber \\
 	\dot{\varphi}_{l} &=& (1+2\alpha)v_{l} - \alpha (v_{l-1}+v_{l+1}) +\psi_{l}-\psi_{l-1} \\
 	\dot{\varphi}_{N}&=& v_{N} - \alpha (v_{N-1}-(1+\gamma)v_{N}) + \psi_{N}-\psi_{N-1}.\nonumber \hspace{0.5cm} 
 	\label{eq:3}
\end{eqnarray} 
where $\gamma=d_s^{0,N}/d_s^l$ is the ratio between the thickness of the first (last) and middle superconducting layers. We assume that due to the proximity effect the thickness of the first and the last S-layer larger than the middle one. Therefore, the nonequilibrium parameters depend on the parameter of boundary conditions $\gamma$, $\eta_{0,N}=\gamma\eta_l $, where $l=1,2,..,N-1$.

\section{Stationary Charge Imbalance}
\begin{figure}[htb]
 \centering
\includegraphics[width=65mm]{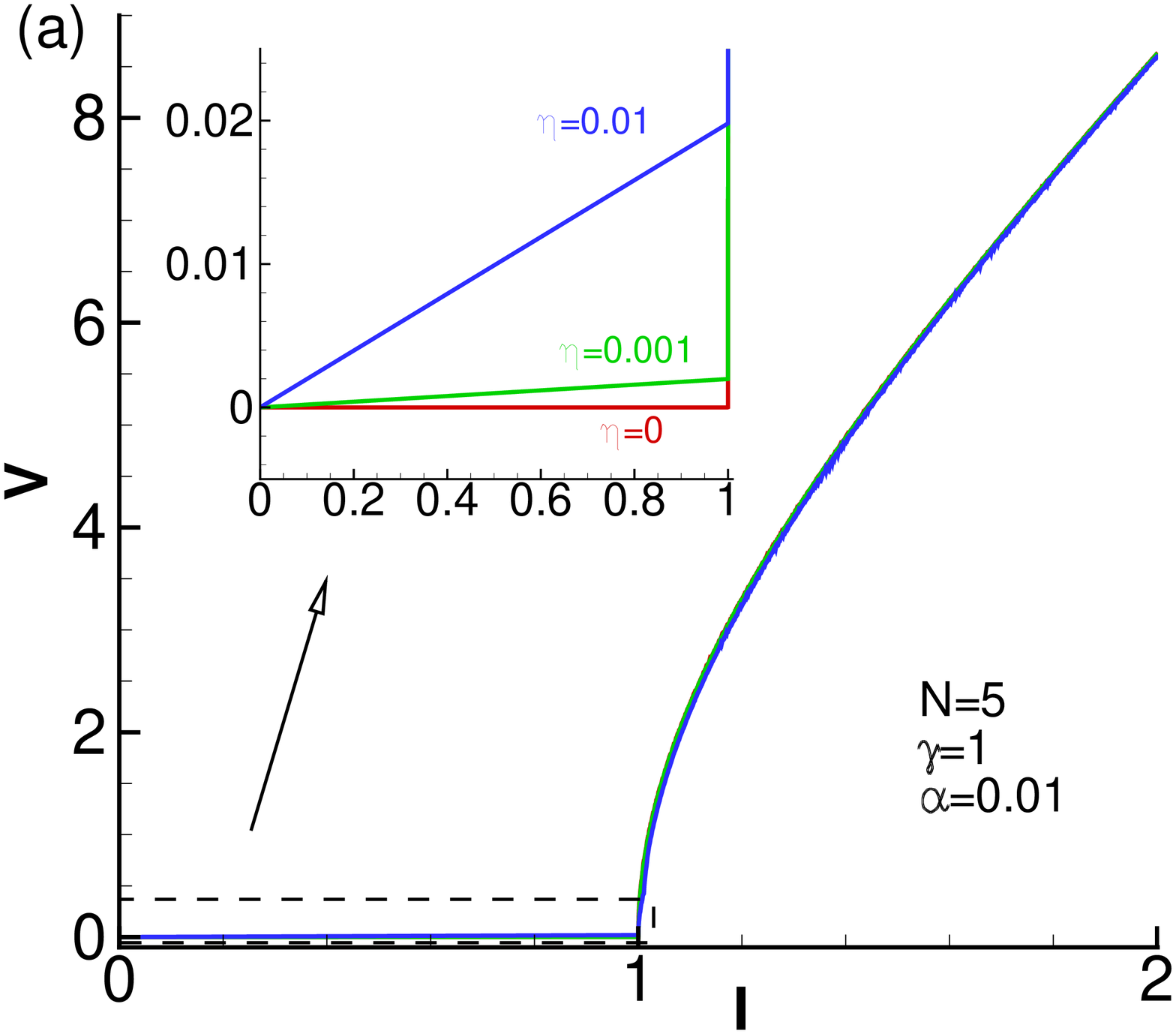}
\includegraphics[width=65mm]{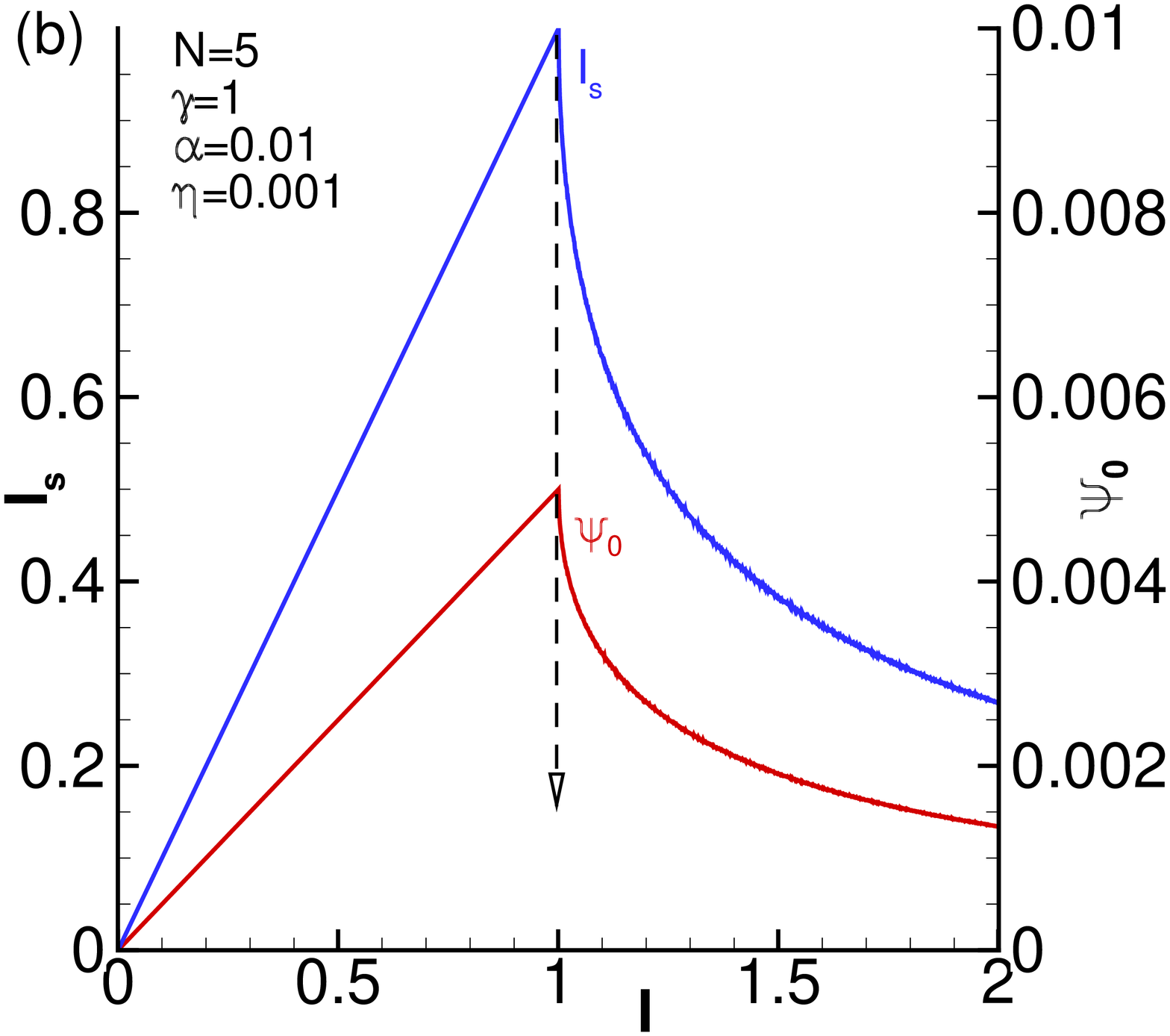}
\caption{(a) IV-characteristics of the system of $N=5$ coupled JJ: (red curve) at $\eta=0$, (green curve) at $\eta=0.001$, (blue curve) at $\eta=0.01$. Inset enlarge zero voltage state on IV-characteristic. (b) Average nonequilibrium potential on layer $0$ together with the average superconducting current at $\eta=0.001$.}
 \label{fig1}
\end{figure}

The main purpose of our paper is to demonstrate a specific influence of stationary charge imbalance on the phase dynamics of the system of coupled JJ. The IV-characteristic together with the average nonequilibrium potential for current bias system of coupled JJs is shown in Fig.\ref{fig1}. IV-characteristic has no hysteresis since there is no capacitance in the system. The inset of Fig.\ref{fig1}(a) enlarge the zero voltage state. The presence of nonequilibrium potential gives a non zero voltage at each current step. So, the finite slope appears on the IV-characteristics in the superconducting state. In Fig.\ref{fig1}(b) the raise of the average nonequilibrium potential of the $0$-th layer $\Psi_0$ in the zero voltage state (left to a dashed line) is shown. Note, that charge imbalance potential has almost negligible value in the S-layers of middle JJs and $\Psi_N$ is symmetrical to $\Psi_0$ with a negative sign. Thus, we limited ourself by showing only $\Psi_0$. It is important to notice that charge imbalance potential has nonzero value in the first S-layer when all JJs are in the R-state. It is possible because $I_s$ has a nonzero value in this region.

\begin{figure}[htb]
 \centering
\includegraphics[width=65mm]{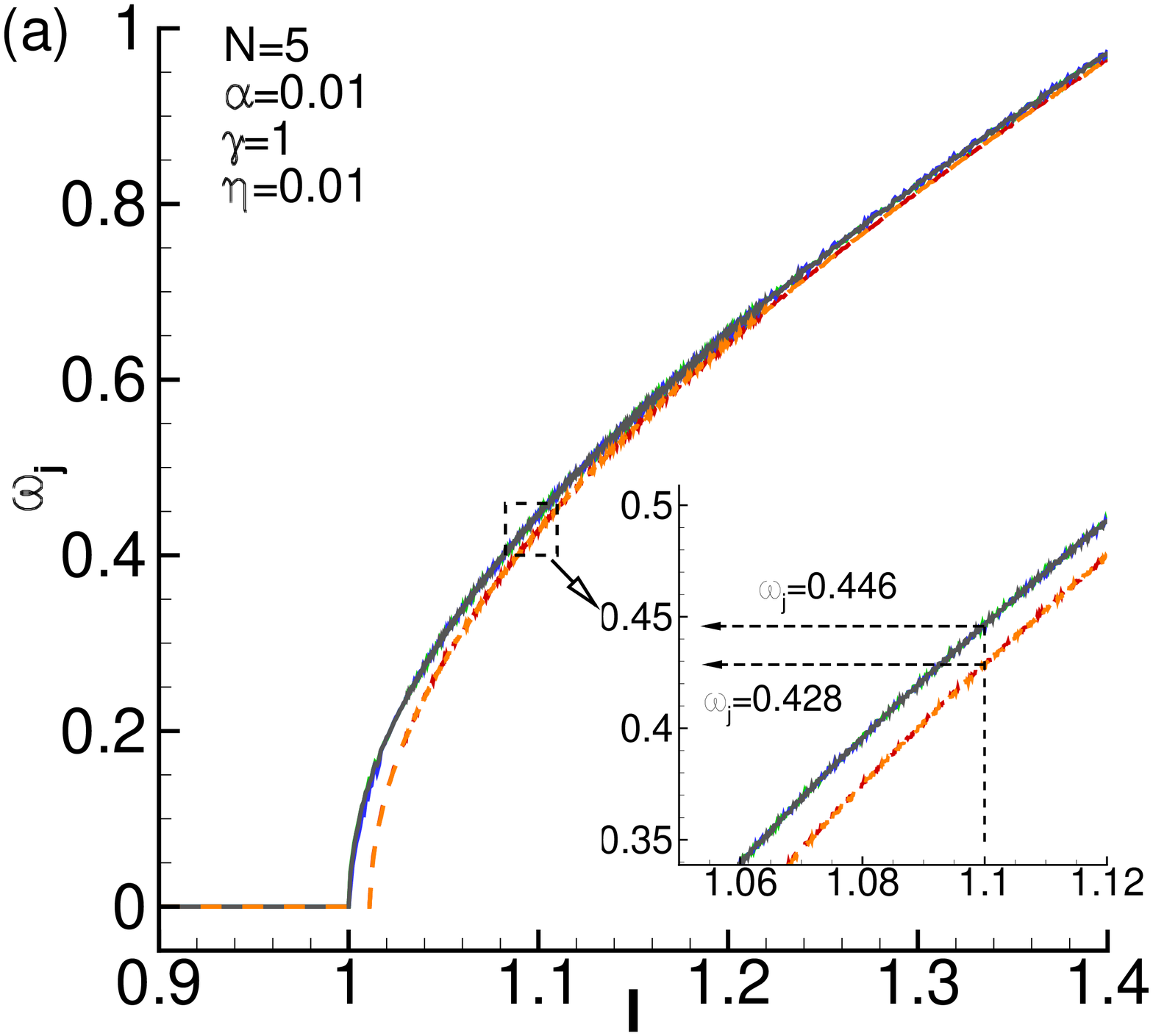}
\includegraphics[width=65mm]{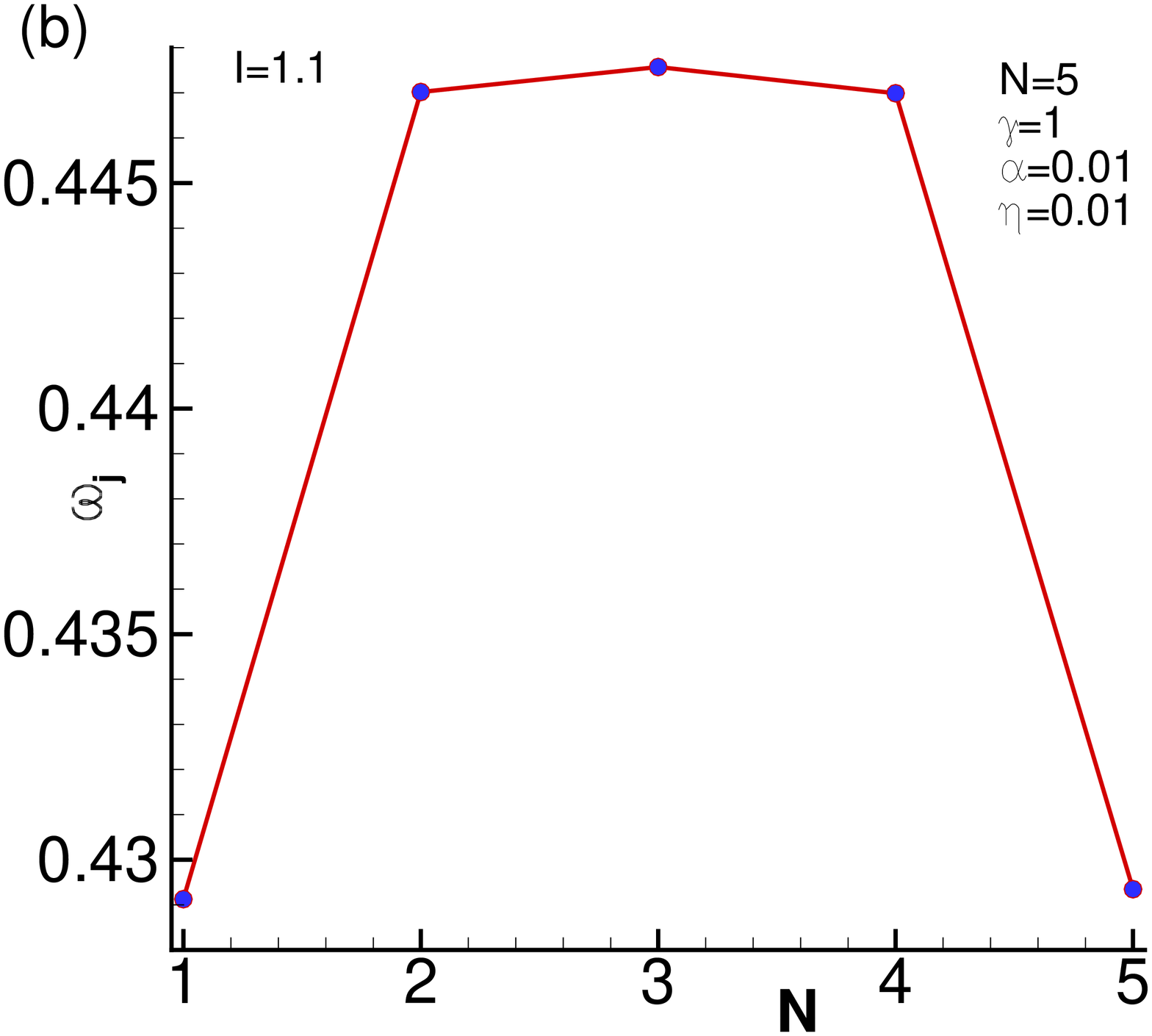}
\caption{(a) Josephson frequency in each JJ as a function of bias current for the system of $N=5$ junctions. Inset enlarge a part indicated by dashed rectangle at current ($I=1.1$). (b) Distribution of Josephson frequency $\omega_j$ along the stack.}
 \label{fig2}
\end{figure}

Important fact here is that a presence of the charge imbalance in the R-state gives a shift of derivative of phase $\frac{d \varphi}{dt}$ in the system of equations (\ref{eq:3}). Notice, that the derivative of phase is equal to the Josephson frequency $\omega_j$. Thus, we show Josephson frequency as a function of current in Fig.\ref{fig2}(a). Dashed curves are $\omega_j$ of the first and last JJs, and solid lines are related to the middle JJs. The inset in Fig.\ref{fig2}(a) shows that for the fixed value of current, the Josephson frequency of the first and last JJs are smaller than in the middle junctions. Therefore, nonequilibrium potential decreases the frequency of oscillations in the JJs on the edges of the stack. Fig.\ref{fig2}(b) demonstrates a distribution of $\omega_j$ along the stack.

\section{Shapiro step at nonequilibrium condition}
\begin{figure}[htb]
 \centering
\includegraphics[width=65mm]{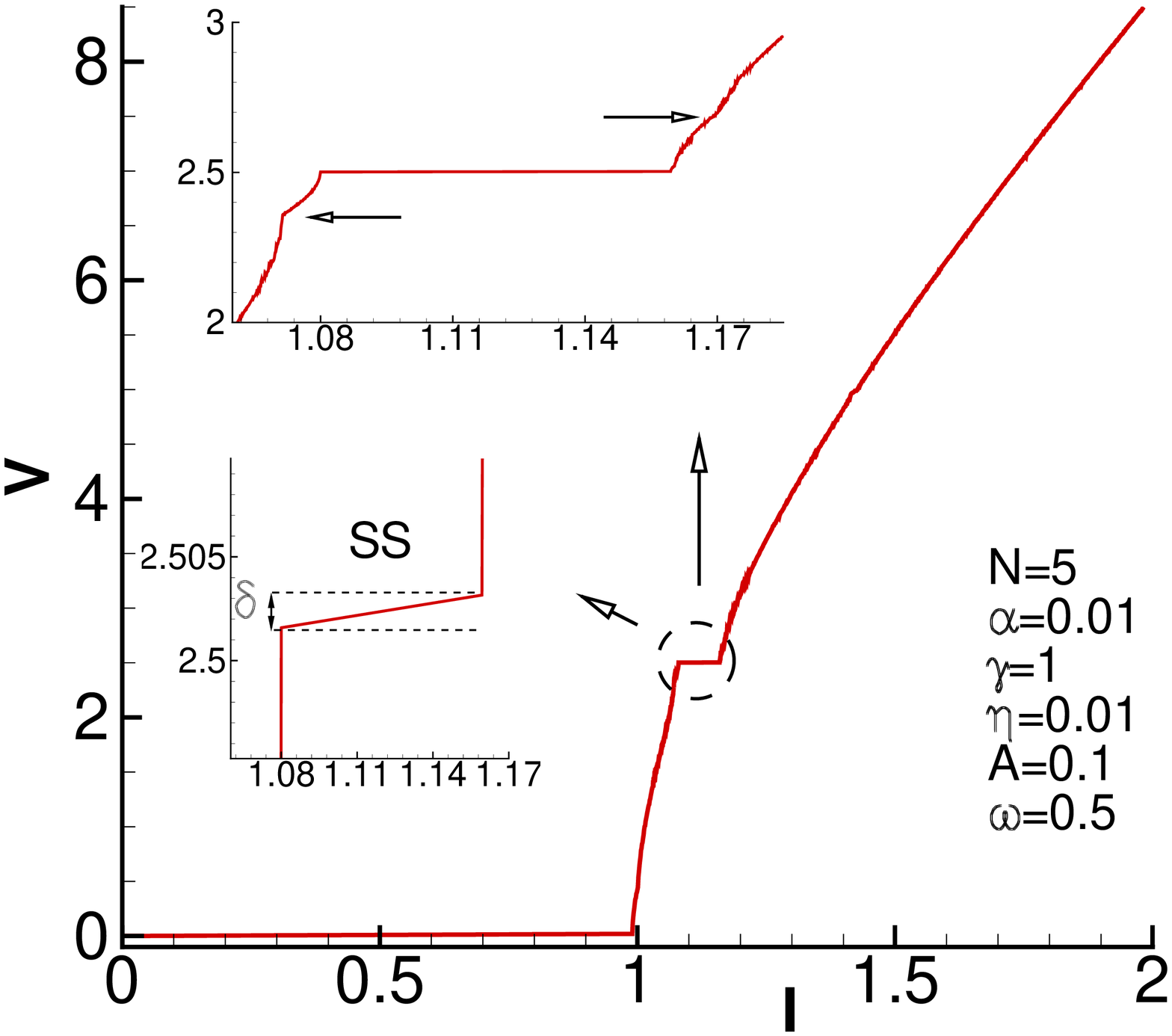}
\caption{IV-characteristic of the system of $N=5$ coupled JJ under external electromagnetic radiation. Lower inset enlarge Shapiro step region to indicate slope. Upper inset enlarge Shapiro step region together with the kinks of voltage below and above Shapiro step indicted by hollow arrows.}
 \label{fig3}
\end{figure}

Let us discuss now the influence of the stationary charge imbalance on the Shapiro step which appears in the IV-characteristics under external radiation. The IV-characteristic calculated for such case is represented in Fig.\ref{fig3}. In the lower inset of Fig.\ref{fig3} the Shapiro step region is enlarged to demonstrate a slope of the step. The slope $\delta$ is marked by two dashed lines. The value $\delta$ depends on the nonequilibrium parameter and has a maximum in the first and last JJs, decreasing in the middle of the stack with decreasing of $\Psi$-potential \cite{epl}. 

\begin{figure}[htb]
 \centering
\includegraphics[width=70mm]{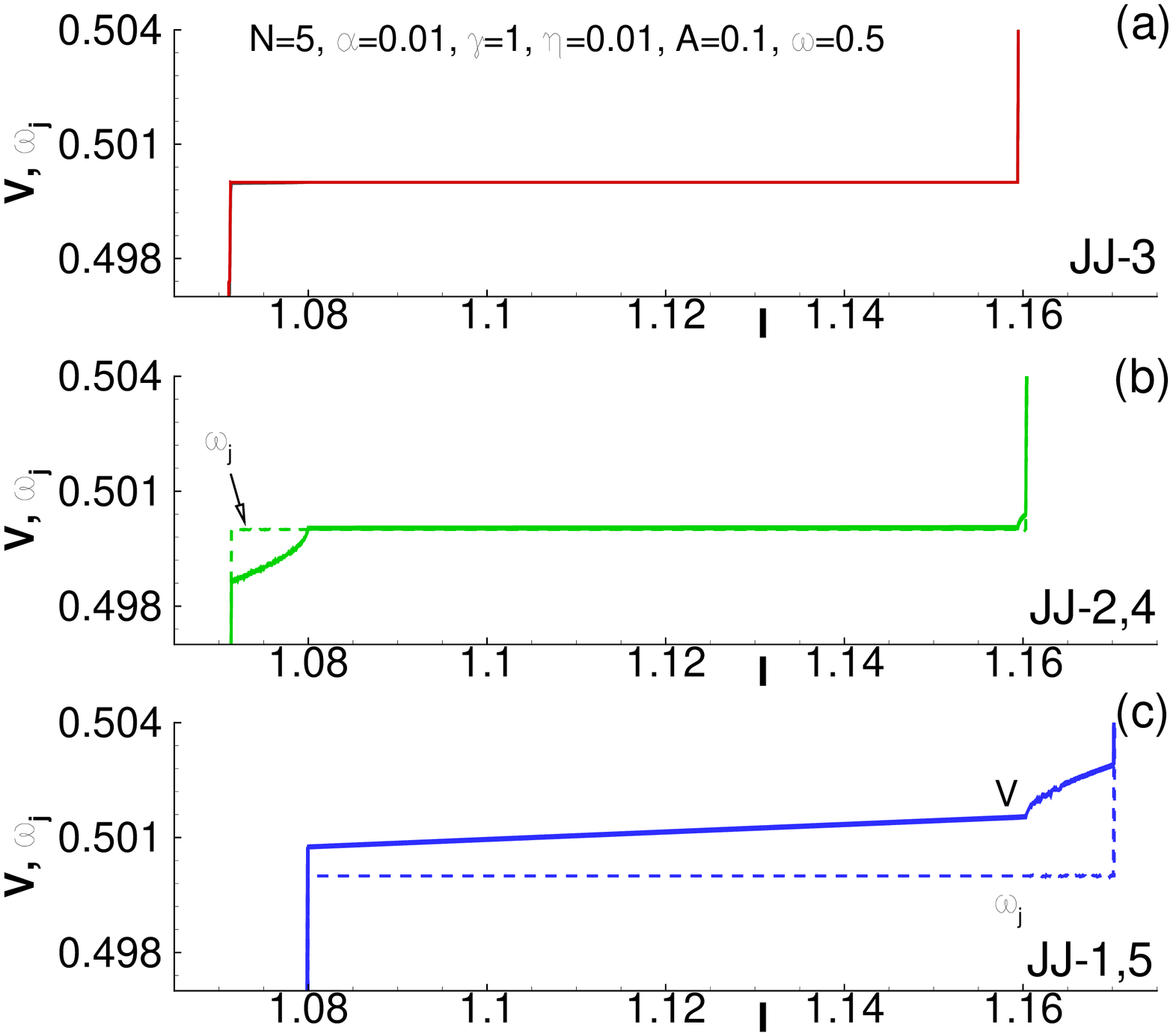}
\caption{Enlarged part of IV-characteristics with Shapiro step (solid lines) together with Josephson frequency as function of current (dashed lines): (a) for third JJ; (b) for second and forth JJ; (c) for first and fifth JJ.}
 \label{fig4}
\end{figure}

The upper inset of Fig.\ref{fig3} demonstrates another interesting effect, namely, small kinks of voltage below and above Shapiro step. Those kinks appear due to the fact that JJs in the stack switched nonsimultaneously to the locked regime. Particularly, junctions in the middle of the stack switched to the locked regime at a smaller current. It brings to the situation when the frequency of the middle junction is equal $\omega_R$ but the frequency of the junctions on edges of the stack is still smaller than $\omega_R$. The details of this process can be seen in Fig.\ref{fig4}. It demonstrates enlarged part of IV-characteristic with Shapiro step (solid lines) together with Josephson frequency as a function of current (dashed lines). Fig.\ref{fig4}(a) demonstrates that third JJ is locking at $I=1.071$ with increasing in bias current. Furthermore, junctions $2$ and $4$ are also in locked regime at this value of current, since $\omega_j$ (see dashed line in Fig.\ref{fig4}(b)) is equal to radiation frequency $\omega_R$ in those junctions at this current. On the other hand, the voltage value is smaller then $V=N\omega_R$ until the value of current $I=1.079$. The cause of this is a coupling of the middle JJs to the first and last ones, which are outside of Shapiro step region (see Fig.\ref{fig4}(c)). The first and last junctions switch to the locked regime at the current $I=1.079$ as it shown in Fig.\ref{fig4}(c). Therefore, the middle part of the stack ``jumps'' to the Shapiro step at a smaller current than JJs on the edges and it gives a small corrections of total voltage. We can see the revers situation above the Shapiro step. Middle JJs ``jump'' off Shapiro step at smaller current when junctions on the edges of the stack are still on the Shapiro step. 

We assume that the observed kinks might be used as basis for development of the method of experimental measurement of coupling value and the value of nonequilibrium parameter, because the distance and the size of those kinks depend on $\eta$ and $\alpha$, and the frequency difference in the JJs of the stack.


\section{Shapiro step on the internal branch}

In the previous theoretical \cite{Ryndy-kkell} and experimental \cite{rother03} works the influence of the stationary charge imbalance on the Shapiro step properties have been investigated in the voltage bias coupled JJs. It was shown that Shapiro step shifts down when the first or last JJ is in the resistive state in comparison with the position of Shapiro step where one of the middle JJs is in the resistive state. The results have been obtained for the overdamped JJs. 

\begin{figure}[htb]
 \centering
\includegraphics[width=65mm]{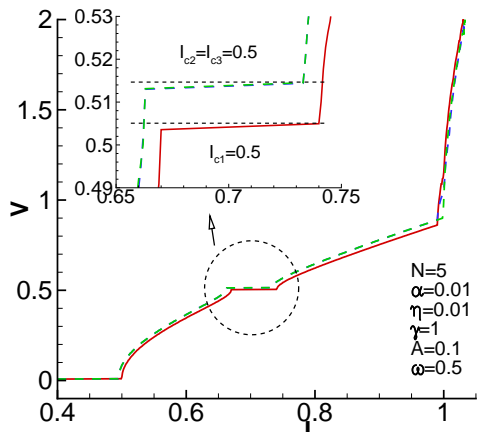}
\caption{Enlarged part of IV-characteristics of the system of $N=5$ coupled JJs with the internal branch and Shapiro step on it: (solid curve) in the case when first JJ has smaller critical current $I_c=0.5$, (dashed curve) in the case when second or third JJ has smaller $I_c=0.5$. Inset enlarge Shapiro step region.}
 \label{fig5}
\end{figure}

In our study we concentrate on the current bias coupled JJs. We model the same situation when one of the JJs has a smaller critical current $I_c=0.5$, in this case one JJ switch to R-state at smaller current and internal branch appears on the IV-characteristic. We investigate the Shapiro step features on the internal branch taking into account stationary charge imbalance effect. In Fig.\ref{fig5} the enlarged part of IV-characteristic of $N=5$ coupled JJ with the obtained internal branch is shown. The inset enlarge Shapiro step region of the IV-characteristic where one of the junction in the resistive state and others are in S-state. The upper curves correspond to the case when one of the middle JJ (second or third) has a smaller critical current. The lower one correspond to the case when first JJ is in R-state. Notice, that Shapiro step shifts down when first JJ is in R-state. This effect is in a perfect agreement with the results obtained in Ref. \cite{Ryndy-kkell, rother03}. 


Furthermore, Shapiro steps demonstrate a finite slope (dashed line is horizontal in the inset) due to the difference of the nonequilibrium potentials on the opposite edges of the step \cite{epl}. We note, that it is impossible to obtained the slope of Shapiro step in case of voltage bias JJs \cite{rother03}. Therefore, we reproduced the experimental results obtained in the case of current bias system of coupled JJ and was able to notice additional manifestation of charge imbalance.    

\section{Summary}

We have investigated the effect of the stationary charge imbalance on the system of coupled JJ. The main result of this study is that nonzero charge imbalance in the R-state leads to a decrease of the Josephson frequency in the JJs of the stack. There are two main reasons due to which it is possible. First, the value of the superconducting current is steel significant in the R-state in the case of overdamped JJs and thus, nonequilibrium potential is not zero as well. Second, that the thickness of the superconductive layers is small enough to have a coupling between the S-layers. The distribution of the $\omega_j$ along the stack strongly depend on the value of nonequilibrium parameter. In our case the $\Psi$-potential decay very fast along the stack. In particular, we have $\Psi_{0,5} \gg \Psi_{2,3,4}$ and noticeable difference in $\omega_j$ has only first and last JJs. The difference in Josephson frequencies leads to the nonuniform switch to the Shapiro step regime in the presence of external radiation and appearance of kinks of voltage on the IV-characteristics of the stack. We assume that our results can be used as a basis for the experimental measurements of the nonequilibrium potential and coupling parameter. Since, distance between those kinks on the IV-characteristics and they size strongly depend on frequency difference of JJ in the stack.   

We also have shown that stationary charge imbalance brings to the slope of the Shapiro step due to the difference of the charge imbalance potential on the edges of the step. This slop have been investigated in details for the system of coupled JJ in the presence of nonstationary charge imbalance in Ref.\cite{epl}.

Finally, we have calculated the IV-characteristics of the current bias system of coupled JJ where one of JJs is in resistive state and have compared the theoretical \cite{} and experimental \cite{rother03} results of voltage bias system of coupled JJ. The difference in Shapiro step shift have been shown for the situation where first JJ has a smaller $I_c$ and where second or third JJ has smaller $I_c$. Additionally, it has been shown that Shapiro step on the internal branch has a finite slope due to the influence of the charge imbalance.    

\section{Acknowledges}
The authors thank I. Rahmonov for fruitful discussions. This work is supported by RFBR grant 18-32-00950.

\addcontentsline{toc}{chapter}{Bibliography}

\begin{thebibliography}{0}



\bibitem{bkg_prb-11} T. M. Benseman, A. E. Koshelev, K. E. Gray, et al., Phys. Rev. B, 84, 064523 (2011).

\bibitem{k_prb-10} A. E. Koshelev, Phys. Rev. B, 82, 174512 (2010).

\bibitem{pasu_prb-08} J. Pfeiffer, A. A. Abdumalikov, Jr., M. Schuster, and A. V. Ustinov, Phys. Rev. B, 77, 024511 (2008).

\bibitem{yurgens2000intrinsic} A. A. Yurgens, Supercond. Sci. Technol., 13, R85 (2000).

\bibitem{okk_sci-07} L. Ozyuzer, A. E. Koshelev, C. Kurter, et al., Science, 318, 1291 (2007). 

\bibitem{Artemenko97} Artemenko S. and Kobelkov A., Phys. Rev. Lett., 78, 3551 (1997)

\bibitem{Preis98} C. Preis, C. Helm, J. Keller, A. Sergeev, \and R. Kleiner, Superconducting Superlattices II: Native and Artificial., I. Bozovic and D. Pavona, editors, Proceedings of SPIE, 3480, 236 (1998).

\bibitem{Shafranjuk99} S. E. Shafranjuk \and M. Tachiki, Phys. Rev. B, 59, 14087 (1999).

\bibitem{Helm00} C. Helm, C. Preis, C. Walter, J. Keller, Phys. Rev. B, 62, 6002 (2000).

\bibitem{Helm01} C. Helm, J. Keller, C. Preis, \and A. Sergeev, Physica C, 362, 43 (2001).

\bibitem{Helm02} C. Helm, L. N. Bulaevskii, E. M. Chudnovsky, M. P. Maley, Phys. Rev. Lett., 89, 057003 (2002).

\bibitem{Bulaevskii02} L. N. Bulaevskii, C. Helm, A. R. Bishop, M. P. Maley, Europhys. Lett., 58, 057003 (2002).

\bibitem{keller05} J. Keller, D. A. Ryndyk, Phys. Rev. B, 71, 054507 (2005).

\bibitem{Koyama96} T. Koyama \and M. Tachiki, Phys. Rev. B, 54, 16183 (1996).

\bibitem{Dmitry} D. A. Ryndyk, Phys. Rev. Lett., 80, 3376 (1998).
	
\bibitem{rother03} S. Rother, Y. Koval, P. M{\"u}ller, R. Kleiner, D.A. Ryndyk, J. Keller, C. Helm, Phys. Rev. B, 67, 024510 (2003).

\bibitem{sm-prl2007} Yu. M. Shukrinov \and F. Mahfouzi, Physica C: Superconductivity, 460, 1303 (2007).

\bibitem{Ryndy-kkell} D. A. Ryndyk, J. Keller \and C. Helm., J. Phys.: Condens. Matter, 14, 815 (2002).

\bibitem{epl} Yu. M. Shukrinov, M. Nashaat, K. V. Kulikov, R. Dawood, H. El Samman \and Th. M. El Sherbini, EPL, 115, 20003 (2016).

\bibitem{ryndyk1997quasiparticle} Ryndyk D.A., Journal of Experimental and Theoretical Physics Letters, 65. 791 (1997)

\bibitem{clarke1972experimental} Clarke J., Physical Review Letters, 28, 1363 (1972)

\bibitem{yu1972electric} Yu M.L. and Mercereau J.E., Physical Review Letters, 28, 1117 (1972).

\bibitem{clarke1974measurements} Clarke J. and Paterson J.L., Journal of Low Temperature Physics, 15, 491 (1974).

\bibitem{av_spu-79} S. N. Artemenko and A. F. Volkov, Sov. Phys. Usp., 22, 295 (1979).

\bibitem{e-ZETF-71}G. M. Eliashberg, Zh. Eksp. Teor. Fiz., 61, 1254 (1971).

\bibitem{SM-PhysC2006} Yu. M. Shukrinov, F. Mahfouzi, \and P. Seidel, Physica C: Superconductivity, 449, 62 (2006).

\bibitem{sm-physc2006} Yu. M. Shukrinov, \and F. Mahfouzi, J. Phys.: Conf. Ser., 43, 1 (2006).






%
%
%
%
%
%
%
%
%
%









\end{thebibliography}

\end{document}